%% file: ronbun.tex
\documentstyle[twocolumn,graphicx]{jpsj}
\newcommand{\eps}{\varepsilon}

\newcommand{\vct}[1]{\mbox{\boldmath #1}}
\def\runtitle{Magnetic and Orbital States and Their Phase Transition 
of the Perovskite-Type Ti Oxides}
\def\runauthor{Masahito {\sc Mochizuki} and Masatoshi {\sc Imada}}
\title{Magnetic and Orbital States and Their
Phase Transition of the Perovskite-Type Ti Oxides:
Strong Coupling Approach}
\author
{Masahito {\sc Mochizuki} and Masatoshi {\sc Imada} }

\inst
{Institute for Solid State Physics, 
University of Tokyo,\\ 5-1-5 Kashiwa-no-ha, Kashiwa, Chiba 277-8581}

\recdate{\today}

\abst{
The properties and mechanism of the magnetic phase transition of the 
perovskite-type Ti oxides, which is 
driven by the Ti-O-Ti bond angle distortion, are studied theoretically
by using the effective spin and pseudospin Hamiltonian
with strong Coulomb repulsion.   
It is shown that the A-type antiferromagnetic (AFM(A)) to 
ferromagnetic (FM) phase transition occurs
as the Ti-O-Ti bond angle is decreased. 
Through this phase transition, the orbital state changes only little
whereas the spin-exchange coupling along the $c$-axis 
is expected to change from positive to negative 
nearly continuously and approaches zero 
at the phase boundary.
The resultant strong two-dimensionality in the spin coupling 
causes rapid suppression 
of the critical temperature, as observed experimentally.
It may induce large quantum fluctuations in this region.}

\kword
{perovskite-type Ti oxides, ${\rm GdFeO}_3$-type distortion, 
$d$-type Jahn-Teller distortion, orbital degrees of freedom, 
orbital ordering, second-order perturbation theory, 
A-type antiferromagnetism, two-dimensional spin coupling,
Mermin and Wagner's theorem}
\begin{document}
\sloppy
\maketitle
\input{sec1.tex}

\input{sec2.tex}
\input{sec3.tex}

\input{sec4.tex}
\input{bib.tex}

\end{document}

%% file: sec1.tex
\section{Introduction}

 Electronic and magnetic properties of perovskite-type 
transition-metal oxides with strong Coulomb correlations 
have recently attracted considerable interest from the
viewpoint of a complex interplay of 
charge, spin and orbital degrees of freedom.
The chemical formula of these compounds is $RM{\rm O}_3$, where
$R$ denotes a trivalent rare-earth ion (i.e., La, Pr, Nd,..., Y) 
and $M$ is a transition-metal ion (i.e., Ti, V,...,Ni, Cu). 
This system is appropriate for a systematic study on
the roles of orbital degrees of freedom since we can 
control a certain kind of lattice parameters as magnitudes
of some structural distortions, which strongly affect 
the one-electron bandwidth, the lifting of
the orbital-level degeneracy and the ways of orbital hybridization.

Theoretically, several single-band models have   
succeeded in explaining several properties.
However, within single-band models, it is hard to explain
such phenomena as 
magnetic orderings accompanied by orbital orderings.
According to the pioneering work of Kugel and Khomskii, it is
important to take account of the $3d$-level degeneracy when we 
consider the magnetic and electronic properties of
these compounds~\cite{Kugel72,Kugel73,Kugel82,Khomskii73}.
They have also pointed out that the magnetic ordering and the orbital 
ordering are closely related.
Usually, ${M}{\rm O}_6$ octahedron in perovskite structure 
undergoes some lattice distortions which work 
as lowering of the symmetry and lifting of the 
level degeneracy. Consequently, the structure yields more complicated 
magnetic properties of these compounds.
 
Perovskite-type Ti oxides $R{\rm TiO}_3$ ($R$ being a trivalent
rare-earth ion) is a typical Mott-Hubbard insulator~\cite{Imada98}.  
${\rm Ti}^{3+}$ has a $3d^1$ configuration, and one of the threefold 
$t_{2g}$ orbitals is occupied at each transition-metal 
site ($t_{2g}^1$ configuration). They have also attracted 
interest since these systems show
various magnetic and orbital ordered phases. 
Moreover, it is required
to take the spin and orbital degrees of freedom into consideration
on an equal footing to explain such rich phases.

\begin{figure}[tdp]
\hfil
\includegraphics[scale=0.3]{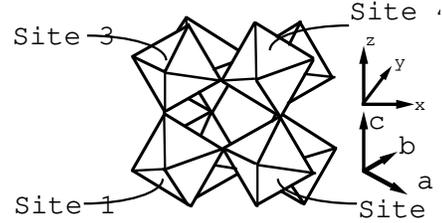}
\hfil
\caption{${\rm GdFeO}_3$-type distortion.}
\label{gdfo3}
\end{figure}

 The crystal structure is an orthorhombically
distorted perovskite (${\rm GdFeO}_3$-type distortion)
in which the ${\rm TiO}_6$ octahedra forming the perovskite 
lattice tilt alternatingly.
In this distortion, the unit cell contains
four octahedra, as shown in Fig.~\ref{gdfo3}.
The magnitude of the distortion (in other words, degree of 
the Ti-O-Ti bond-angle distortion) depends on the ionic radii
of the $R$ ion. With a small ionic radius of the $R$ ion, 
the lattice structure is more distorted and the bond angle
is decreased more significantly from $180^{\circ}$.
For example, in ${\rm LaTiO}_3$, the bond angle is 
$157^{\circ}$ ($ab$-plane)
and $156^{\circ}$ ($c$-axis), but $144^{\circ}$ ($ab$-plane) 
and $140^{\circ}$ ($c$-axis) in ${\rm YTiO}_3$~\cite{MacLean79}.
We can control the magnitude or the Ti-O-Ti bond angle
by the use of solid-solution systems  
${\rm La}_{1-y}{\rm Y}_{y}{\rm TiO}_3$
or in $R{\rm TiO}_3$, by varying the $R$ ions
from La, Pr, Nd,..., to Y.
In particular, by varying the Y concentration in 
${\rm La}_{1-y}{\rm Y}_{y}{\rm TiO}_3$, we can control the 
bond angle almost continuously from $156^{\circ}$
($y = 0$) to $140^{\circ}$ ($y = 1$).
Mainly, the distortion has two kind of roles.
One is the ``band-width control''.
Since the transfers of electrons on Ti $3d$ orbitals are governed by 
the super-transfer processes mediated by the O $2p$ states in 
perovskite-type transition-metal oxides, the Ti-O-Ti bond angle distortion 
causes the reduction of $3d$ $t_{2g}$-electron bandwidth $W$
critically.
Another role of the Ti-O-Ti bond angle distortion is a symmetry relaxation
of the indirect $d$-$d$ transfer between 
neighboring $3d$ orbitals with different symmetries.
In the cubic-perovskite lattice with no distortion, an electron
in an $t_{2g}$ orbital with a certain symmetry can transfer to that with the 
same symmetry at the neighboring site.
On the other hand, the ${\rm GdFeO}_3$-type distortion relaxes 
the symmetry restriction and makes it possible for an electron 
in a $t_{2g}$ orbital with one symmetry to transfer to that with another 
symmetry at the neighboring site.
In addition, the indirect transfers between neighboring 
$t_{2g}$ and $e_g$ orbitals are increased critically 
as the distortion increases. 
This is one of the reasons for a rich magnetic phase diagram
as a function of the bond angle
since the magnetic and orbital orderings are strongly associated 
with the way of orbital-hybridization as was studied previously by
Kanamori~\cite{Kanamori59,Kanamori60},
Goodenough~\cite{Goodenough55,Goodenough63}, 
Kugel and Khomskii~\cite{Kugel72,Kugel73,Kugel82,Khomskii73}.
In this paper, ${\rm GdFeO}_3$-type distortion 
is simulated by rotating the ${\rm TiO}_6$ octahedra
by angle $+{\theta}$ and $-{\theta}$ about the $(1,1,1)$ and $(-1,-1,1)$
axes with respect to the $x$, $y$, and $z$ axes.

 In ${\rm YTiO}_3$, a $d$-type Jahn-Teller (JT) 
distortion has been observed in which the
longer and shorter Ti-O bond lengths are 
$\sim$2.08 $\AA$ and $\sim$2.02 $\AA$, respectively~\cite{Akimitsu98}.
On the other hand, ${\rm LaTiO}_3$ exhibits a small or no JT
distortion.
In the $d$-type JT distortion, the elongated axes of
the octahedra are parallel along the $c$-axis. 
In the JT distortion, antibonding character between 
Ti $3d$ and O $2p$ orbitals is reduced
and the orbitals directed in the elongated direction
are lowered in energy.
For example, if the $x$-axis is elongated, among the threefold
$t_{2g}$ orbitals, the $xy$ and
$zx$ are lowered relative to the $yz$.
As a result, in the $d$-type JT distortion 
the $xy$ and $yz$ orbitals are stabilized
at sites 1 and 3 and the $xy$ and $zx$ orbitals are stabilized
at sites 2 and 4. 
The JT distortion plays an important role
on orbital and magnetic orderings. 

\begin{figure}[tdp]
\hfil
\includegraphics[scale=0.3]{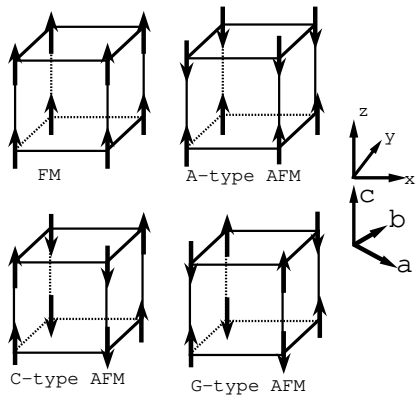}
\hfil
\caption{Typical magnetic structures for the perovskites}
\label{sp-struct}
\end{figure}
 Recently, electronic and magnetic phase diagrams have been 
investigated intensively as functions of the magnitude of
a Ti-O-Ti bond angle distortion
~\cite{Goral82,Greedan85,Okimoto95,Katsufuji97}.
In the less-distorted or in a La-rich($y<0.6$)
region, the system shows an AFM ground state.
In particular, ${\rm LaTiO}_3$ ($y=0.0$) shows a G-type AFM(AFM(G)) 
ground state with a magnetic moment of 0.45 $\mu_{\rm B}$, 
in which the spins are aligned antiferromagnetically
in all directions~\cite{Goral83}.
With increasing Y concentration or varying the $R$ site with 
smaller-sized ions (a decrease of the Ti-O-Ti bond angle),
the N${\rm {\grave{e}}el}$ temperature ($T_N$)
decreases rapidly and is suppressed to almost zero,
subsequently a FM ordering appears.
This rapid decrease of $T_N$ is not well explained
by the conventional models. Its origin is 
an issue of interest.
In the relatively distorted or Y-rich region, the system has
a FM ground state. In ${\rm YTiO}_3$ ($y=1.0$), the value of
the magnetic moment is 0.84 $\mu_{\rm B}$
and the Curie temparature ($T_C$) takes $\sim 30$ K~\cite{Garret81}.
This ferromagnetism is hardly explained by a simple single-band Hubbard
model and requires consideration of the $d$-level degeneracy.

Strong suppressions of $T_N$ and $T_C$ around the AFM-FM transition
point imply a continuous-type transition at $T=0$.
At first sight, this is a puzzling feature, because we expect
the first-order transition between completely different 
symmetry breaking at $T=0$ and $T_N$ and $T_C$ may remain 
nonzero at the transition point.
Clarifying the mechanism of this continuous-type transition 
is the purpose of our study in this paper.
  
Recent model Hartree-Fock(HF) studies based 
on a multiband $d$-$p$ model have succeeded in reproducing
the spin structures of both end compounds, ${\rm LaTiO}_3$ and 
${\rm YTiO}_3$~\cite{Mizokawa96a,Mizokawa96b}.
According to the studies, 
the $t_{2g}^1$ configuration in ${\rm LaTiO}_3$
with smaller ${\rm GdFeO}_3$-type distortion and with small or no
JT distortion is well described by the spin-orbit ground state,
out of which two states with antiparallel orbital and spin moments,
$\frac{1}{\sqrt{2}}(z^{\prime}x^{\prime}+iy^{\prime}z^{\prime})\uparrow$ and
$\frac{1}{\sqrt{2}}(z^{\prime}x^{\prime}-iy^{\prime}z^{\prime})\downarrow$,
with the $z^{\prime}$-axis pointing in the (1,1,1)-direction
in terms of the $x$, $y$ and $z$ axes are alternating
between nearest neighbors, favored both by the spin-orbit interaction
and by the super-exchange interaction.
As a result, the AFM(G) state in which
the spins point in the $z^{\prime}$-direction is 
expected to be realized.
In addition, the total energies of various spin and orbital structures
are calculated as functions of the Ti-O-Ti bond angle in the large
$d$-type JT distortion.
As a result, in the large ${\rm GdFeO}_3$-type distortion, a FM solution
accompanied by an orbital ordering was proved to be stabilized. 
Moreover, the band calculation for ${\rm YTiO}_3$ by 
the generalized gradient approximation (GGA) and 
the local spin-density-approximation (LSDA) in which the
${\rm GdFeO}_3$-type distortion is taken into account also succeeded
in reproducing a FM spin state with an orbital ordering
in ${\rm YTiO}_3$~\cite{Sawada97}.

In these weak coupling approach, however, the 
$d$-$d$ Coulomb interaction is
treated in an approximate and averaged way while
in perovskite-type Mn oxides, the importance of the interaction 
in the degenerate orbitals has been pointed out 
in terms of the magnetic and orbital ordering by
Kanamori~\cite{Kanamori59,Kanamori60}, 
Goodenough~\cite{Goodenough55,Goodenough63}, 
Kugel and Khomskii~\cite{Kugel72,Kugel73,Kugel82,Khomskii73}.
Actually, the intrasite Coulomb interaction is also much
larger than the other leading energies of the parameters in 
Ti systems.
The GGA and LSDA are also not sufficient for the description
of the strongly localized electron states.
It is well-known that the GGA and LSDA have a tendency to 
underestimate the magnitude of the band gaps.
Although the FM state is obtained both in LSDA and GGA
for ${\rm YTiO}_3$, that obtained in LSDA is metallic and 
that obtained in GGA is half-metallic, being in 
disagreement with the experimental result that 
${\rm YTiO}_3$ is insulating with optical gap of 1.0 eV~\cite{Okimoto95}.
Moreover, any magnetic solutions cannot be obtained for 
${\rm LaTiO}_3$ in LSDA.
These facts indicate that the electron correlation
should be treated in more sophisticated way when we
consider the electronic properties of these compounds.

Moreover, the magnetic and orbital states realized in 
the actual systems and the properties of their
phase transitions are still controversial.
If the JT distortion is small,
the spin-orbit interaction is substantial for
the electronic states of the Ti 3d electrons and hence, 
the system is well described by the spin-orbit ground state.
On the other hand, in the region of large JT distortion, the 
energy splitting due to the JT distortion becomes
comparable or larger than the spin-orbit level-splitting.
The situation is no longer the same as that 
in the region of small JT distortion.
The AFM phase realized in such a large JT distortion
can be qualitatively different from that realized
in ${\rm LaTiO}_3$. 
Thus the magnetic and orbital states
realized in the moderately distorted region between
${\rm LaTiO_3}$ and ${\rm YTiO_3}$
are issues of interest.
In addition, the nature of the magnetic phase transition,
which is considered to occur in the large
JT-distortion-region remains insufficiently clarified.

 In this paper, we study the magnetic and the orbital orderings and
their phase transitions in perovskite-type Ti oxides
as functions of a Ti-O-Ti bond-angle distortion
by using an effective spin and pseudospin Hamiltonian 
constructed through the second-order perturbational expansion
with respect to the transfer terms in the limit of the strong Coulomb
repulsion.
In this Hamiltonian, the full degeneracy of Ti $3d$ orbitals
and on-site Coulomb and exchange interactions are taken into account.
In addition, effects of ${\rm GdFeO}_3$-type and JT
distortions are also considered by modifications of the
hopping integrals and splitting of the
$t_{2g}$ and $e_g$ levels.
Our approach is appropriate for
the systematic study on the properties and 
mechanism of the magnetic and orbital phase transitions
since the origin of the stabilization of magnetic and 
orbital ordered states are attributed to the second-order perturbational 
energy gains with respect to the indirect $d$-$d$ transfers 
and, moreover, their phase transition are caused by the competition
of their energy gains.
These energy gains are easily investigated in our approach 
by estimating the anisotropic transfer-amplitudes 
and the level-splitting energies which are driven 
by the lattice structure and several lattice distortions.  

We show that:
\begin{itemize}
\item The $e_g$ orbital degrees of freedom play important roles
on the magnetic phase transition in this system;
\item as the ${\rm GdFeO}_3$-type distortion increases, the spin-exchange
interaction along the $c$-axis changes from positive to negative
due to the super-exchange processes
mediated by the $e_g$ orbitals, but the orbital state
does not change;
\item the AFM-FM phase transition point is well described by the
two-dimensional Heisenberg model so that $T_N$ and $T_C$ 
are suppressed to almost zero;
\item large quantum fluctuations and anisotropy in the spin-wave dispersion
are expected to be observed. 
\end{itemize}

 The organization of this paper is as follows.
In Sec. 2, we explain how to construct the effective spin and pseudospin
Hamiltonian of perovskite-type Ti oxides.
In Sec. 3, numerical results calculated by utilizing 
a mean field approximation are presented.
Sec. 4 is devoted to the summary and conclusions.
A short version of this paper has been 
published~\cite{Mochizuki00a}, but
this paper contains additional and more detailed results.
In addition, in the previous paper, we simulated the ${\rm GdFeO}_3$-type
distortion by rotating the ${\rm TiO}_6$ octahedra around the
axes in the $ab$-plane, but in this paper, the distortion is
simulated in a more realistic way.

%% file: sec2.tex
\section{Formalism}

\begin{figure}[tdp]
\hfil
\includegraphics[scale=0.5]{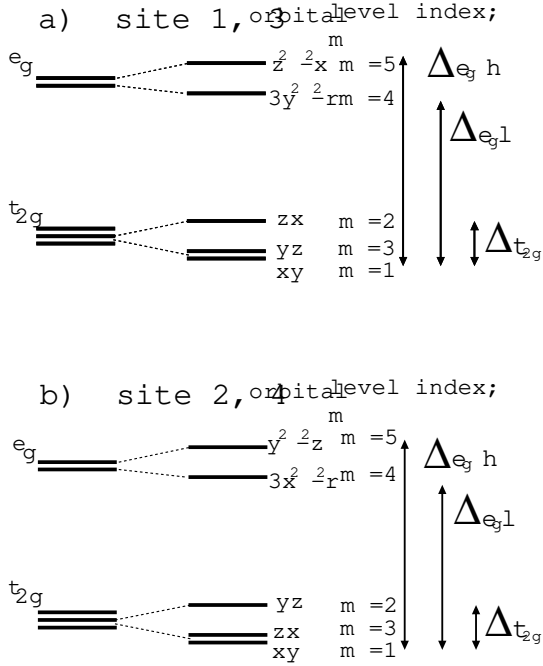}
\hfil
\caption{$3d$-level splitting in the $d$-type JT distortion:
The ways of the level splittings are different between
sites 1, 3 and sites 2, 4.$\quad$Fig. a) shows that in sites 1 and 3
and $\quad$ Fig. b) shows that in sites 2 and 4.}
\label{jtsplit}
\end{figure}

We start with the multiband $d$-$p$ model in which 
the full degeneracies of Ti $3d$ and O $2p$
orbitals as well as the on-site Coulomb and exchange 
interactions are taken into account.
The Hamiltonian is given by
\begin{equation}
        H^{dp} = H_{d0} + H_{p} + H_{tdp} + H_{tpp}
               + H_{\rm on-site} , \\
\label{dph}
\end{equation}
with
\begin{eqnarray}
     & &H_{d0} = \sum_{i,\gamma,\sigma} \eps_{d}^0
        d_{i,\gamma,\sigma}^{\dagger} d_{i,\gamma,\sigma}, \\
     & &H_{p} = \sum_{j,l,\sigma} \eps_{p}
        p_{j,l,\sigma}^{\dagger} p_{j,l,\sigma}, \\
     & &H_{tdp} = \sum_{i,\gamma,j,l,\sigma} 
        t_{i\gamma,jl}^{dp}
                      d_{i,\gamma,\sigma}^{\dagger} 
                      p_{j,l,\sigma}  + \vct{h.c.}, \\
     & &H_{tpp} = \sum_{j,l,j',l',\sigma} 
        t_{jl,j'l'}^{pp}
                      p_{j,l,\sigma}^{\dagger} 
                      p_{j',l',\sigma}  + \vct{h.c.}, \\
     & &H_{\rm on-site} = H_{u} + H_{u'} + 
                            H_j + H_{j'}, 
\label{dp hamiltonian2}
\end{eqnarray} 
where ${d_{i,\gamma,{\sigma}}^{\dagger}}$ 
is a creation operator of an electron 
with spin $\sigma(={\uparrow},{\downarrow})$ in the
$3d$ orbital $\gamma$ at Ti site $i$ and ${p_{j,l,{\sigma}}^{\dagger}}$ 
is a creation operator of an electron 
with spin $\sigma(={\uparrow},{\downarrow})$ in the
$2p$ orbital $l$ at oxygen site $j$.
Here, we choose the representation of the
fourfold symmetry in the $d$-type JT distortion
as the basis of $3d$ orbitals at each site, namely, $xy$, $yz$, $zx$,
$3y^2-r^2$ and $z^2-x^2$ at site 1 and site 3 and 
$xy$, $yz$, $zx$, $3x^2-r^2$ and $y^2-z^2$ at site 2 
and site 4 (see Fig.~\ref{jtsplit}).
$H_{d0}$ and $H_p$ stand for the bare level energy of Ti $3d$ and  
O $2p$ orbitals, respectively.
$H_{tdp}$ and $H_{tpp}$ are $d$-$p$ and $p$-$p$ hybridization
term, respectively.
The term $H_{\rm on-site}$ represents on-site $d$-$d$ Coulomb interactions.
${t_{i\gamma,jl}}^{dp}$ and
${t_{jl,{j^{\prime}}{l^{\prime}}}}^{pp}$ 
are nearest-neighbor $d$-$p$ transfer and $p$-$p$ transfer given in terms of 
Slater-Koster parameters $V_{pd{\pi}}$, $V_{pd{\sigma}}$, $V_{pp{\pi}}$ 
and $V_{pp{\sigma}}$. 
$H_{\rm on-site}$ term consists of the following four contributions:
\begin{eqnarray}
     & &   H_{u} = \sum_{i,m} u
        d_{i,m,\uparrow}^{\dagger} d_{i,m,\uparrow}
        d_{i,m,\downarrow}^{\dagger} d_{i,m,\downarrow}, \\
     & &   H_{u'} = \sum_{i,m>m',{\sigma},{\sigma}'} u'
        d_{i,m,\sigma}^{\dagger} d_{i,m,\sigma}
        d_{i,m',{\sigma}'}^{\dagger} d_{i,m',{\sigma}'}, \\
     & &   H_{j} = \sum_{i,m>m'\sigma,{\sigma}'} j
        d_{i,m,\sigma}^{\dagger} d_{i,m',\sigma}
        d_{i,m',{\sigma}'}^{\dagger} d_{i,m,{\sigma}'}, \\
     & &   H_{j'} = \sum_{i,m \ne m'} j'
        d_{i,m,\uparrow}^{\dagger} d_{i,m',\uparrow}
        d_{i,m,\downarrow}^{\dagger} d_{i,m',\downarrow}, 
\end{eqnarray}
where $H_{u}$ and $H_{u'}$ are the intra- and inter-orbital 
Coulomb interactions and $H_{j}$ and $H_{j'}$ denote the
exchange interactions.
The term $H_{j}$ is the origin of the Hund's rule coupling 
which strongly favors the spin alignment in the same
direction on the same atoms.
The term $H_{j'}$ gives the $\uparrow\downarrow$-pair
transfer between the $3d$-orbitals on the same Ti atom.
These interactions are expressed by using Kanamori parameters,
$u$, $u^{\prime}$, $j$ and $j^{\prime}$ which 
satisfy the following relations~\cite{Brandow77,Kanamori63};
\begin{eqnarray}
     u &=& U + \frac{20}{9}j , \\
     u'&=& u -2j , \\
     j &=& j'.
\end{eqnarray}
Here, $U$ gives a magnitude of the multiplet-averaged 
$d$-$d$ Coulomb interaction.
The charge-transfer energy $\Delta$, which describes the energy
difference between occupied O $2p$ and unoccupied
Ti $3d$ levels, is
defined by using $U$ and energies of the
bare Ti $3d$ and O $2p$ orbitals $\eps_d^0$ and $\eps_p$
as follows,
\begin{equation}
      \Delta = \eps_{d}^0 + U -\eps_p,
\end{equation}
since the characteristic unoccupied $3d$ level energy on the
singly occupied Ti site is $\eps_{d}^0 + U$.
The values of $\Delta$, $U$ and $V_{pd\sigma}$ are
estimated by the cluster-model analyses of valence-band and
transition-metal $2p$ core-level photoemission spectra
~\cite{Saitoh95,Bocquet96}.
We take the values of these parameters as
$\Delta = 7.0$ eV, $U = 4.0$ eV, $V_{pd\sigma} = -2.2$ eV
and $j = 0.64$ eV
throughout the present calculation. 
The ratio $V_{pd\sigma}/V_{pd\pi}$ is fixed at $-2.16$, and
$V_{pp\sigma}$ and $V_{pp\pi}$ at 0.60 eV and $-0.15$ eV, 
respectively~\cite{Harrison89}.
The effects of the ${\rm GdFeO}_3$-type distortion 
are considered through the 
$d$-$p$ transfer integrals which is defined by using the
Slater-Koster's parameters~\cite{Slater54}.
The effects of the $d$-type JT distortion are also
considered. 
The magnitude of the distortion can be denoted
by the ratio $[V_{pd{\sigma}}^s$/$V_{pd{\sigma}}^l]^{1/3}$; 
here, $V_{pd{\sigma}}^s$ and $V_{pd{\sigma}}^l$
are the transfer integrals for the shorter and longer
Ti-O bonds. 
The value for ${\rm YTiO}_3$ estimated using Harrison's
rule~\cite{Harrison89} takes $\sim$1.036.
In order to reveal the nature of the magnetic phase transition 
and the origin of the rapid suppression of $T_N$, 
we focus on the situation near the phase boundary between
AFM and FM phases.
The value of ratio 
$[V_{pd\sigma}^s/V_{pd\sigma}^l]^{1/3}$ is fixed at 
1.030, which is expected to be realized near the phase boundary
under the assumption of linear decrease as a function 
of the bond angle from 1.036 (${\rm YTiO}_3$) to 1.00 (${\rm LaTiO}_3$).
Under the JT distortion, the $t_{2g}$ level-splitting energy
$\Delta_{t_{2g}}$ is estimated to be 0.050 eV.
Since the $t_{2g}$ level-splitting due to the spin-orbit 
interaction is sufficiently small in comparison with $\Delta_{t_{2g}}$,
we neglect the spin-orbit interaction 
through the present calculations. 

In the path-integral formalism, the expression of
the partition function is given by
\begin{equation}
Z = \int {\cal D} {\bar{d}}_{i,\gamma,\sigma}
           {\cal D} d_{i,\gamma,\sigma} 
           {\cal D} {\bar{p}}_{j,l,\sigma}
           {\cal D} p_{j,l,\sigma} 
\exp \left[- \int_0^{\beta} {\rm d}\tau L(\tau) \right] ,
\end{equation}
with
\begin{eqnarray}
         L(\tau ) = H^{dp}(\tau)
& & + \sum_{i,\gamma,\sigma} {\bar{d}}_{i,\gamma,\sigma}(\partial_\tau-\mu)
d_{i,\gamma,\sigma} \nonumber \\       
& & +  \sum_{j,l,\sigma} {\bar{p}}_{j,l,\sigma}
(\partial_\tau-\mu)p_{j,l,\sigma},  
\end{eqnarray}
where $\tau$ denotes the imaginary time introduced in the
path-integral formalism and  
${\bar{d}}_{i,\gamma,\sigma}$, $d_{i,\gamma,\sigma}$,
${\bar{p}}_{j,l,\sigma}$ and $p_{j,l,\sigma}$ 
are the Grassman-variables corresponding to 
the operators  $d_{i,\gamma,\sigma}^{\dagger}$, $d_{i,\gamma,\sigma}$,
$p_{j,l,\sigma}^{\dagger}$ and $p_{j,l,\sigma}$, respectively.
By using the Matsubara-frequency representation:
\begin{eqnarray}               
     d_{i,\gamma,\sigma}(\tau) = \frac{1}{\sqrt{\beta}}
\sum_{\omega_n} d_{i,\gamma,\sigma}(\omega_n) e^{-i\omega_{n}\tau}, \\
     p_{j,l,\sigma}(\tau) = \frac{1}{\sqrt{\beta}}
\sum_{\omega_n} p_{j,l,\sigma}(\omega_n) e^{-i\omega_{n}\tau},      
\end{eqnarray}
we have
\begin{eqnarray}
Z\! =\!\! \int& & {\cal D} {\bar{d}}_{i,\gamma,\sigma}(\omega_n)
           {\cal D} d_{i,\gamma,\sigma}(\omega_n) 
           {\cal D} {\bar{p}}_{j,l,\sigma(\omega_n)}
           {\cal D} p_{j,l,\sigma}(\omega_n) \nonumber \\
    & &\times\exp \left[- \sum_{\omega_n}  L(\omega_n) \right]  
\end{eqnarray}
with
\begin{eqnarray}
         L(\omega_n) &=& \sum_{i,\gamma,\sigma} 
{\bar{d}}_{i,\gamma,\sigma}(-i\omega_n+\eps_d^0-\mu)
d_{i,\gamma,\sigma} \nonumber \\      
& &  + \sum_{j,l,\sigma} 
{\bar{p}}_{j,l,\sigma}(-i\omega_n+\eps_p-\mu)
p_{j,l,\sigma} \nonumber \\
& &  + \sum_{i,\gamma,j,l,\sigma}t_{im,jl}^{dp} 
{\bar{d}}_{i,\gamma,\sigma}p_{j,l,\sigma}  + \vct{c.c.} \nonumber \\
& &  +  \sum_{j,l,j',l'\sigma}t_{jl,j'l'}^{pp}
{\bar{p}}_{j,l,\sigma}p_{j',l',\sigma}  + \vct{c.c.} \nonumber \\
& &  + H_{\rm on-site}(\omega_n). 
\end{eqnarray}
After integrating over the ${\bar{p}}$ and $p$, the partition
function is rewritten as
\begin{equation}
Z = \int {\cal D} {\bar{d}}_{i,\gamma,\sigma}(\omega_n)
           {\cal D} d_{i,\gamma,\sigma}(\omega_n)
\exp \left[- \sum_{\omega_n} L_d(\omega_n) \right] , 
\end{equation}
where
\begin{eqnarray}
L_d(\omega_n) = \sum_{i,\gamma,\sigma}& & 
{\bar{d}}_{i,\gamma,\sigma}(-i\omega_n+\eps_d^0-\mu)d_{i,\gamma,\sigma}  
\nonumber \\   
+\sum_{\omega_n}\sum_{i,\gamma,i',\gamma',\sigma}
\sum_{j,l,j',l'}\!\!&&
\!{\bar{d}}_{i,\gamma,\sigma}
\!\left[ t_{i\gamma,jl}^{dp}\!(H^{-1}_{jl,j'l'}(i\omega_n))
t_{i'\gamma',j'l'}^{dp} \right]  
d_{i',\gamma',\sigma}   \nonumber \\   
& &+H_{\rm on-site}(\omega_n). 
\end{eqnarray}
Here, a matrix $H_{jl,j'l'}(i\omega_n)$ takes the form
\begin{equation}
H_{jl,j'l'}(i\omega_n) = -(-i\omega_n+(\eps_p-\mu)) \delta_{jl;j'l'} \\
         - t_{jl,j'l'}^{pp}.
\end{equation}

Substituting $i\omega_n$ in $H^{-1}_{jl,j'l'}(i\omega_n)$
with characteristic energies of a $3d$ electron,
we can obtain the expressions of the effective $d$-$d$ transfers 
and $3d$ level energies
as follows;
\begin{eqnarray}
    t_{i\gamma,i'\gamma'}^{dd}\!\! &=&\!\! \left\{
\begin{array}{ll}

\sum_{j,l,j',l'}H^{-1}_{jl,j'l'}(\eps_{d}^0+u-\mu) 
t_{i\gamma,jl}^{dp}t_{i'\gamma',j'l'}^{dp} \\
\quad\quad\quad\quad\quad\quad\quad\quad\quad\quad
\mbox{for}\quad\mbox{Case}\quad a)\\
\\
\sum_{j,l,j',l'}H^{-1}_{jl,j'l'}(\eps_{d}^0+u'-\mu)
t_{i\gamma,jl}^{dp}t_{i'\gamma',j'l'}^{dp} \\
\quad\quad\quad\quad\quad\quad\quad\quad\quad\quad
\mbox{for}\quad\mbox{Case}\quad b)\quad , \\
\end{array} \right. 
\\
  \eps_{d\,i,\gamma} &=& \eps_{d}^0 + \sum_{j,l,j',l'}H^{-1}_{jl,j'l'}
(\eps_{d}^0+U-\mu)
t_{i\gamma,jl}^{dp}t_{i\gamma,j'l'}^{dp} .
\end{eqnarray}
\begin{figure}[tdp]
\hfil
\includegraphics[scale=0.5]{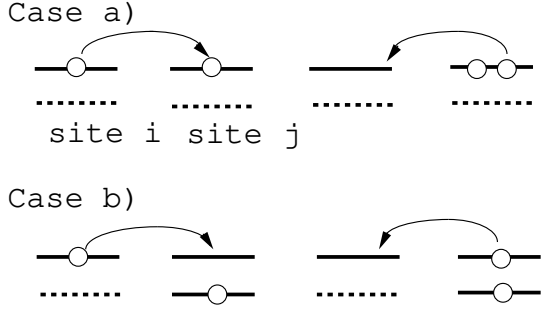}
\hfil 
\caption{Characteristic perturbational 
processes of a $3d$ electron in two examples.
Case a) contains the intermediate states in which 
an orbital occupied by two electrons 
with opposite spins in the process.
Case b) contains only the states in which an orbital 
occupied by one electron.
The energy takes $\eps_{d}^0+u$ in Case a) 
and $\eps_{d}^0+u^{\prime}$ in Case b) at the intermediate states.}  
\label{trprocess}
\end{figure} 
The case a) denotes the processes which contain 
the doubly-occupied-orbital states
before or after the electron transfer to a neighboring site.
On the other hand, the case b) denotes the processes with
no doubly-occupied-orbital states (see Fig.~\ref{trprocess}).
As a result, the "effective" multiband Hubbard Hamiltonian
derived from the multiband $d$-$p$ model has the form;
\begin{equation}
        H^{\rm mH} = H_{d}^{\rm mH} + H_{tdd}^{\rm mH} 
+ H_{\rm on-site},  \\
\end{equation}
with
\begin{eqnarray}
     & &H_{d}^{\rm mH} = \sum_{i,m,\sigma} \eps_{d\,i,m}
        d_{i,m,\sigma}^{\dagger} d_{i,m,\sigma}, \\
     & &H_{tdd}^{\rm mH} = \sum_{i,m,i',m',\sigma} 
        t_{im,i'm'}^{dd}
                      d_{i,m,\sigma}^{\dagger} 
                      d_{i',m',\sigma}  + \vct{h.c.}, \\
     & &H_{\rm on-site} = H_{u} + H_{u'} + 
                            H_j + H_{j'}, 
\label{multihubhamilt}
\end{eqnarray} 
By $H_{d}^{\rm mH}$, we express the level energies of Ti $3d$ orbitals 
under the influence of the crystal fields in the $d$-type 
JT distortion with
\begin{equation}
      \eps_{d\,i,m} = \left\{
      \begin{array}{ll}
        
        \eps_{dl}                      & \mbox{for} \quad m = 1, 3, \\
        \eps_{dl} + \Delta_{t_{2g}}    & \mbox{for} \quad m = 2,  \\
        \eps_{dl} + \Delta_{{e_g}l}   & \mbox{for} \quad m = 4,  \\
        \eps_{dl} + \Delta_{{e_g}h}  & \mbox{for} \quad m = 5 .\\
      \end{array} \right. 
\end{equation}
Here, $m=1,3$ are lower ${t_{2g}}$ levels,
$m=2$ is a higher ${t_{2g}}$ level and $m=4$ and $m=5$ are
lower and higher $e_g$ levels, respectively.
The $\Delta_{t_{2g}}$, $\Delta_{{e_g}h}$ and $\Delta_{{e_g}l}$   
denote the level-splitting energies measured from
lower $t_{2g}$ level as shown in Fig.~\ref{jtsplit}.
It should be noted that the same indices of energy levels 
at different sites do not necessarily correspond to the orbitals with 
the same symmetry.
$H_{tdd}^{\rm mH}$ is a $d$-$d$ super-transfer term. 

Among the energy parameters in the multiband Hubbard 
Hamiltonian, 
the on-site Coulomb interactions have the largest energy scale.
Since the electron filling of the present compounds is
kept at $d^1$ configuration on average, the on-site
Coulomb interactions drive the compounds to the Mott insulating state
with suppressed charge fluctuations.
Therefore, starting with the multiband Hubbard Hamiltonian, 
we can derive an effective Hamiltonian in the low-energy region
on the subspace of states
only with singly occupied $t_{2g}$ orbitals
at each transition-metal site by utilizing a 
second-order perturbation theory.
The states of $3d$ electron localized at the transition-metal
sites can be represented by two quantum numbers, the 
$z$-component of the spin $S_z$
and the number of the occupied orbitals. 
When one of the twofold lower $t_{2g}$ orbitals
is occupied at each site, we can describe the electronic states 
using a spin-1/2 operators, which we call the pseudospin $\vct{$\tau$}$. 
We can describe the occupied energy-level 1 
by a quantum number ${\tau}_z = -1/2$,
and level 3 by ${\tau}_z = +1/2$. 
We follow an approach similar to the well-known
Kugel-Khomskii formulation~\cite{Kugel72,Kugel73,Kugel82,Khomskii73}.
We express the $3d$ electron operators in terms of
$\vct{S}$ and $\vct{$\tau$}$ to arrive at 
the effective spin and pseudospin Hamiltonian:
\begin{equation}
      H_{\rm eff} = \tilde {H}_{d}^{\rm mH} + H_{t_{2g}} + H_{e_g},  
\label{eqn:eqhamlt}    
\end{equation}
where
\begin{equation}
      \tilde {H}_{d}^{\rm mH} = \sum_{i,\sigma,m,=1,2,3} \eps_{d\,i,m}
        d_{i,m,\sigma}^{\dagger} d_{i,m,\sigma}. \\
\end{equation}
The first term $\tilde{H}_{d}^{\rm mH}$
is obtained from the zeroth-order perturbational processes.
The second term $H_{t_{2g}}$ is
obtained from the second-order perturbational processes whose 
intermediate states contain only $t_{2g}$-orbital degrees
of freedom.
The third term $H_{e_g}$ is obtained 
from the second-order perturbational processes whose 
intermediate states contain $e_g$-orbital degrees
of freedom.
In this Hamiltonian, the exchange interactions between 
neighboring spins and orbitals are characterized by the 
energies in the intermediate states in the perturbational processes.
The spin configuration in the system is determined by the
competition of the perturbational energy gains which depend on the
orbital states and amplitudes of the
anisotropic transfer-integrals. 
Moreover, in this Hamiltonian, the spin and orbital 
configurations are not determined independently. 
The terms $H_{t_{2g}}$ and $H_{e_g}$ 
are described by products of spin and pseudospin operators
since the second-order perturbational processes simultaneously
change both spin and orbital states in adjacent sites.
In this sense, spin and orbital degrees of freedom 
strongly couple with each other.

In this Hamiltonian, the $e_g$-orbital degrees of freedom are
taken into account as virtual states of perturbational 
processes.
It is considered that if the energy splitting between 
$t_{2g}$ and $e_g$ levels ($\Delta_{e_g}$)
is sufficiently large relative
to that due to the JT distortion or a characteristic energy scale
of low-energy excitations, the relevant electronic orbitals
for low-energy excitations or ground-state properties are 
the $t_{2g}$ orbitals, and the $e_g$-orbital degrees of freedom are
negligible.
However, in perovskite-type Ti oxides, this energy splitting
is rather small relative to the other perovskite-type transition-metal
oxides since the charge-transfer energy is relatively large.
We can approximately estimate the difference $\Delta_{e_g}$
as follows,
\begin{equation}
  \Delta_{e_g} \sim 3\frac{V_{pd\sigma}^2}{\Delta}
                  - 4\frac{V_{pd\pi}^2}{\Delta},
\label{eqn:eqdleg}
\end{equation}
where $V_{pd\sigma}$ and $V_{pd\pi}$ are the Slater-Koster
parameters for the $p$-$d$ transfer. 
The expression of Eq. (\ref{eqn:eqdleg}) can 
be easily obtained by Slater and Koster's 
relations~\cite{Harrison89,Slater54}.
In $R{\rm TiO}_3$ case, this value takes about 1.4 eV. 
The order of energy reduction
due to the virtual transfer mediated by a singly-occupied $t_{2g}$  
state is given by $\frac{t^2}{u} = \frac{t^2}{(u'+2j)}$
and that mediated by an unoccupied $e_g$ state is  
$\frac{t^2}{(u'+\Delta_{e_g})}$.
Since the values of $\Delta_{e_g}$ and $2j$ are
comparable in $R{\rm TiO}_3$, 
$\quad\frac{t^2}{(u'+\Delta_{e_g})}$-terms
are not negligible compared to the $\frac{t^2}{u}$-terms.
Hence, the $e_g$-orbital degrees of freedom cannot be neglected
even in the low-energy region so that 
we take them into consideration as virtual states of the perturbational
processes.
In general, the charge-transfer energy $\Delta$ gradually
increases as the atomic number of the transition-metal decreases.

In addition, we have also examined the effective Hamiltonian
on the subspace of states in which an electron
occupies not only twofold lower $t_{2g}$ levels
but also a higher $t_{2g}$ level in the JT distortion 
at each site by replacing with a pseudospin 
representation whose magnitude
takes $1$.
However, in the large JT distortion as 
$[V_{pd{\sigma}}^s$/$V_{pd{\sigma}}^l]^{1/3}=1.030$
in the present case,  
the occupancy of the higher $t_{2g}$ level is close to zero
so that the results are not changed from those
obtained by using the Hamiltonian Eq. (\ref{eqn:eqhamlt}).

%% file: sec3.tex
\section{Numerical Results and Discussions}

In this section, we present the numerical results calculated by 
applying a mean-field approximation 
to the effective spin and pseudospin Hamiltonian
introduced in the previous section.
We have introduced the following averages
$\langle S_{\gamma_1}\rangle$, $\langle \tau_{\gamma_1}\rangle$ and 
$\langle S_{\gamma_1}\tau_{\gamma_2}\rangle$ 
with $\gamma_1, \gamma_2 = x, y$ and $z$
as the mean fields at each site 
in the ${\rm GdFeO}_3$-type unit cell, which are to be determined 
self-consistently.

We have calculated the total energies of various 
spin and orbital configurations as functions of the bond angle.
Without the ${\rm GdFeO}_3$-type distortion
or in the small distortion 
region ($\angle$Ti-O-Ti$= 180^{\circ}\sim151^{\circ}$),
FM solution with ($yz,xy,xy,zx$)-type orbital ordering
in which site 1, 2, 3 and 4 are dominantly occupied by 
$yz, xy, xy$, and $zx$, respectively (FM1 solution)
is stabilized. We can specify the orbital state realized in the 
FM1 solution by using two angles $\theta_1$ and $\theta_2$ 
as follows, 
\begin{eqnarray}
&{\rm site}& 1; \quad\cos{\theta_1}|yz>+\sin{\theta_1}|xy>,
\nonumber \\
&{\rm site}& 2; \quad\cos{\theta_2}|zx>+\sin{\theta_2}|xy>,
\nonumber \\
&{\rm site}& 3; \quad -\cos{\theta_2}|yz>+\sin{\theta_2}|xy>,
\nonumber \\
&{\rm site}& 4; \quad -\cos{\theta_1}|zx>+\sin{\theta_1}|xy>.
\label{eqn:eqtheta1}
\end{eqnarray}

\begin{figure}[tdp]  
\hfil
\includegraphics[scale=0.5]{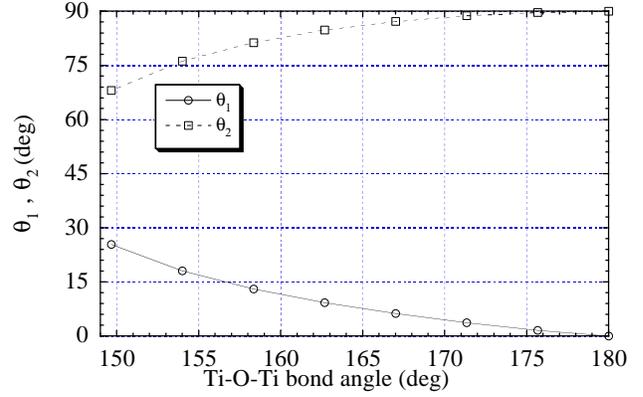}
\hfil
\caption{The orbital structure in the FM1 solution
as a function of the Ti-O-Ti bond angle.
Without the ${\rm GdFeO}_3$-type distortion, a complete
($yz,xy,xy,zx$)-type orbital order is realized.}
\label{theta1}
\end{figure}

\begin{figure}[tdp]
\hfil
\includegraphics[scale=0.5]{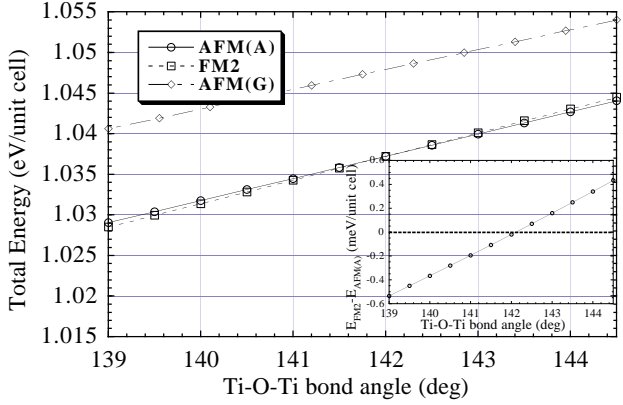}
\hfil
\caption{Total energies of AFM(A), FM2 and AFM(G)
solutions near the AFM-FM phase boundary are
plotted as functions of the Ti-O-Ti bond angle.
Inset shows the energy difference between the AFM(A) and FM2
solutions.}
\label{rel_ene}
\end{figure}

In Fig.~\ref{theta1}, the angles $\theta_1$ and $\theta_2$ are plotted 
as functions of the Ti-O-Ti bond angle.
Without the ${\rm GdFeO_3}$-type distortion ($\angle$Ti-O-Ti$=180^{\circ}$),
complete ($yz, xy, xy, zx$)-type occupation is realized.
With increasing the ${\rm GdFeO_3}$-type distortion, 
the occupations of the $xy$, $zx$, $yz$ and $xy$
orbitals gradually increase at site 1, 2, 3 and 4, respectively.
In this orbital ordering, the neighboring occupied orbitals are 
approximately orthogonal and electron-transfers from the occupied
orbitals are restricted to neighboring unoccupied orbitals.
This spin and orbital configuration is favored both by 
transfers and by the exchange interaction $j$ in the
small ${\rm GdFeO}_3$-type distortion.
However, it should be noted that 
in our study, the JT-distortion parameter
is fixed at a large value
in order to focus on the situation realized
near the phase boundary.
On the contrary, in ${\rm LaTiO}_3$ or in the less distorted region,
there is no JT distortion and spin-orbit ground state is considered
to be realized.
So that, the AFM(G) state is not reproduced in this calculation
even in the less distorted region.
In addition, we should note that the FM1 solution which is stabilized
in small ${\rm GdFeO}_3$-type and large JT distortions is not realized
in the actual systems. 
By moderately decreasing the Ti-O-Ti bond angle
(increasing the ${\rm GdFeO}_3$-type distortion),
the A-type AFM(AFM(A)) solution with another type of
orbital ordering is stabilized rather than
the FM1 solution around $\angle$Ti-O-Ti $\sim 151^{\circ}$.
With further decreasing of the Ti-O-Ti bond angle, a FM state
accompanied by an orbital ordering is stabilized again.
The type of the orbital ordering in this FM phase
is similar to that in the AFM(A) phase.
Hereafter, we refer to this FM state as FM2.
The AFM to FM phase transition observed in the actual system
corresponds to this AFM(A) to FM2 phase transition.
In Fig.~\ref{rel_ene}, total energies of the AFM(A), FM2 and 
AFM(G) states are plotted near the phase boundary.
The AFM(G) solution has much higher energy relative to other
two solutions. The AFM(A) to FM2 phase transition occurs
at $\angle$Ti-O-Ti$\sim142^{\circ}$.

We can specify the orbital states realized in the AFM(A) and FM2
solutions by using the angle $\theta_{\rm AFM(A)}$ and 
 $\theta_{\rm FM2}$ as,
\begin{eqnarray}
&{\rm site}& 1; \quad\cos{\theta_x}|xy>+\sin{\theta_x}|yz>,
\nonumber \\
&{\rm site}& 2; \quad\cos{\theta_x}|xy>+\sin{\theta_x}|zx>,
\nonumber \\
&{\rm site}& 3; \quad -\cos{\theta_x}|xy>+\sin{\theta_x}|yz>,
\nonumber \\
&{\rm site}& 4; \quad -\cos{\theta_x}|xy>+\sin{\theta_x}|zx>,
\end{eqnarray}
where $x =$ AFM(A), FM2.
\begin{figure}[tdp]
\hfil
\includegraphics[scale=0.5]{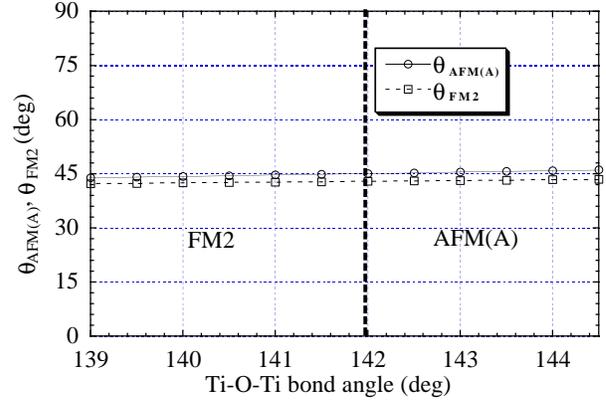}
\hfil
\caption{The orbital structures in the AFM(A) and FM2 solutions 
as functions of the Ti-O-Ti bond angle.
The difference between those of the two solutions 
are considerably small.}
\label{theta2}
\end{figure}
In Fig.~\ref{theta2}, the angles for the AFM(A) and FM2 solutions 
($\theta_{\rm AFM(A)},\theta_{\rm FM2}$) are plotted.
The difference between the $\theta_{\rm AFM(A)}$ and $\theta_{\rm FM2}$
is very small and both take almost the same value ($\sim 45^{\circ}$).
This indicates that the way of the orbital ordering 
changes only little through the magnetic phase transition.
Hereafter, we refer to the orbital state realized in the AFM(A)
and FM2 phases as ($yz,zx,yz,zx$)-type orbital order.

\begin{figure}[tdp]
\hfil
\includegraphics[scale=0.5]{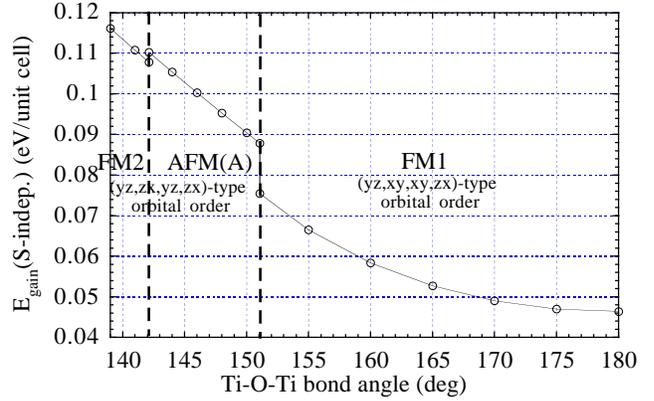}
\hfil
\caption{The absolute values of spin-independent energy gains
for stabilized solutions as functions of the Ti-O-Ti 
bond angle, which are calculated by substituting 
$S$ with 0.}
\label{enegS0}
\end{figure}
\begin{figure}[tdp]
\hfil
\includegraphics[scale=0.5]{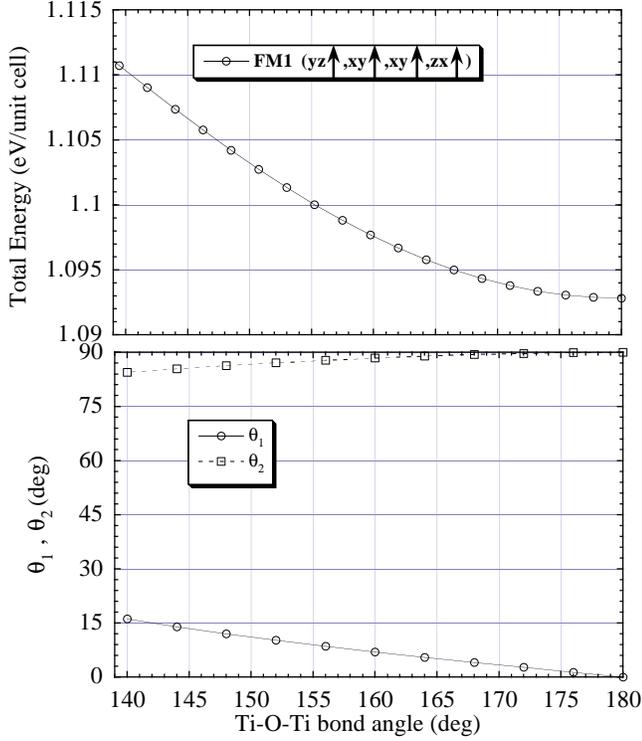}
\hfil
\caption{(Upper panel); The total energy for the FM1 solution calculated 
by utilizing a Hamiltonian which includes
only $t_{2g}$ orbital degrees of freedom
($H_{\rm eff} = \tilde{H}_{d}^{\rm mH} + H_{t_{2g}}$).
Within this model, the other spin and orbital configurations
do not have any stable solutions.
(Lower panel); The orbital structures in this FM1 solution.}
\label{enewfweg}
\end{figure}

In Fig.~\ref{enegS0}, the absolute values
of the spin-independent energy gains due to the
second-order perturbations ($E_{\rm gain}$($S$-indep.))
are plotted for the stabilized spin and orbital structures.
The figure shows that ($yz,xy,xy,zx$)-type to ($yz,zx,yz,zx$)-type
orbital state phase transition 
accompanied by a large jump of the energy-gain  
occurs at $151^{\circ}$ and the value increases rapidly 
as the bond angle decreases in the ($yz,zx,yz,zx$)-type 
orbital order region. 
This indicates that the ($yz,zx,yz,zx$)-type orbital order
is strongly favored as Ti-O-Ti bond angle is decreased.
As the ${\rm GdFeO}_3$-type distortion increases,
the indirect hybridization between the 
neighboring $t_{2g}$ orbitals and $e_g$ orbitals increases. 
Since the hybridizations between neighboring $e_g$ and O $2p$
orbitals have a $\sigma$-bonding character, amplitudes of the
transfers between these orbitals are critically increased 
by the distortion.
Hence, the $e_g$-orbital degrees of freedom
becomes indispensable for the stability of the ($yz,zx,yz,zx$)-type
orbital ordering and the orbital ordering is 
strongly stabilized with increasing the ${\rm GdFeO}_3$-type 
distortion because of the large amplitudes of transfers
toward neighboring $e_g$ orbitals.
Actually, within the model which does not contain the $e_g$-orbital
degrees of freedom, only the FM solution with ($yz,xy,xy,zx$)-type 
orbital order (FM1) is stabilized and the ($yz,zx,yz,zx$)-type orbital
state does not have any stable solutions.
Moreover, the energy of the FM1 state increases as the magnitude of
the ${\rm GdFeO}_3$-type distortion is increased as shown in 
Fig.~\ref{enewfweg} (upper panel).
This tendency is consistent with the fact that $t_{2g}$ 
bandwidth is reduced as the Ti-O-Ti bond angle decreases.
The orbital state realized in this FM1 solution
can also be specified by utilizing two angles $\theta_1$ and $\theta_2$
as Eq. (\ref{eqn:eqtheta1}). The angles $\theta_1$ and $\theta_2$ are plotted
in Fig.~\ref{enewfweg} (lower panel).

\begin{figure}[tdp]
\hfil
\includegraphics[scale=0.5]{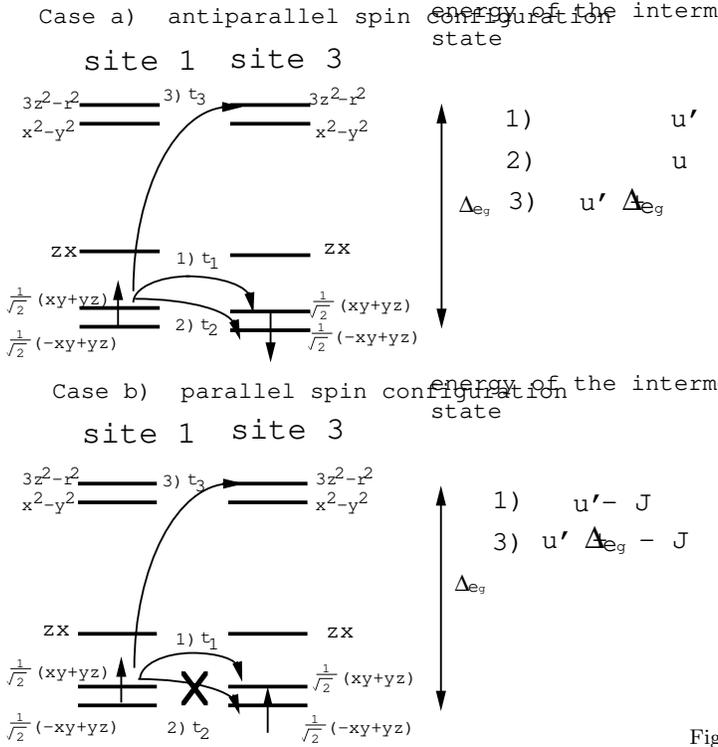}
\hfil
\caption{The second-order perturbational energy gain
strongly depends on the spin configuration.
Substantial transfers along the $c$-axis for the energy gain are 
illustrated for both the parallel and antiparallel spin 
configurations. The angles ${\theta}_{\rm AFM(A)}$ and 
${\theta}_{\rm FM2}$ are fixed at $45^{\circ}$.
The cross symbol $\times$ represents the forbidden transfer.}
\label{sdproc}
\end{figure}

\begin{figure}[tdp]
\hfil
\includegraphics[scale=0.45]{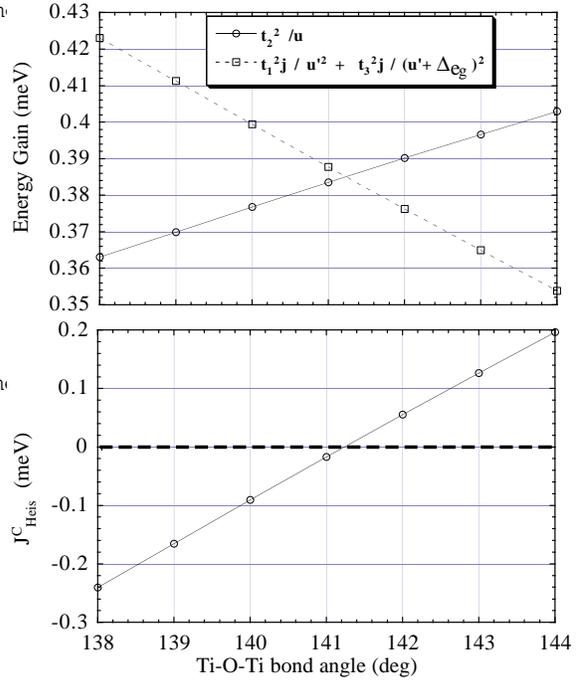}
\hfil
\caption{A characteristic second-order perturbational 
energy gain due to the transfers along the $c$-axis
for the antiparallel spin configuration $t_2^2/u$ and
that for the parallel spin configuration
$t_1^2j/u'^2+ t_3^2j/(u'+\Delta_{e_g})^2$ are plotted as 
functions of the Ti-O-Ti bond angle.
Two energies are crossing at about  $141^{\circ}$.
With decreasing the 
Ti-O-Ti bond angle, the sign of the spin-exchange
interaction along the $c$-axis changes from positive
to negative continuously at the phase boundary.}
\label{egain_Jconbi}
\end{figure}

The AFM(A)-to-FM2 phase transition is identified 
as the transition where the sign of the
spin-exchange interaction along the $c$-axis changes
from negative to positive while that in the $ab$-plane
is constantly negative.
The constant FM coupling in the $ab$-plane
under the ($yz,zx,yz,zx$)-type orbital state can be easily
understood.
In the $ab$-plane, the neighboring orbitals are 
approximately orthogonal to each other.
Hence, the FM spin configuration is favored 
through Hund's rule coupling interaction.
However, the emergence of the FM2 phase is 
not understood straightforwardly 
since the neighboring orbitals along the $c$-axis
are not orthogonal.
By considering the transfers from an occupied orbital
to neighboring $e_g$ orbitals,
this is understood schematically as follows.
At this stage, based on the fact that the orbital state
hardly changes between two phases, we fix the angle 
${\theta}_{\rm AFM(A)}$ and ${\theta}_{\rm FM2}$ at $45^{\circ}$.
Namely, we assume that an electron occupies 
$\frac{1}{\sqrt{2}}(xy+yz)$, $\frac{1}{\sqrt{2}}(xy+zx)$,
$\frac{1}{\sqrt{2}}(-xy+yz)$ and $\frac{1}{\sqrt{2}}(-xy+zx)$ 
at sites 1, 2, 3, and 4, respectively through the phase transition.
In addition, we use the representation of the cubic symmetry
$x^2-y^2$ and $3z^2-r^2$ for the $e_g$ orbitals for intuitive understanding.
Let us consider the energy gain of an electron in the 
$\frac{1}{\sqrt{2}}(xy+yz)$ orbital at site 1, which 
is caused by the second-order perturbational
processes with respect to the transfers along the $c$-axis 
(i.e., the transfers between site 1 and site 3). 
In the large ${\rm GdFeO}_3$-type distortion, the 
$\frac{1}{\sqrt{2}}(xy+yz)$ orbital
at site 1 mainly hybridizes with the $yz$ and $3z^2-r^2$ orbitals
at site 3 along the $z$ direction relative to the other
orbitals. 
For example, the transfer amplitudes between $yz$ at site 1 and
$zx$ or $xy$ orbitals at site 3 are smaller by as much as $10^{-2}$
than those between $yz$ at site 1 and $yz$ or $3z^2-r^2$ orbitals at site 3.
When the $\frac{1}{\sqrt{2}}(-xy+yz)$
orbital at site 3 is occupied by an electron,
the second-order perturbational energy gain of an electron
in the $\frac{1}{\sqrt{2}}(xy+yz)$ 
orbital at site 1 depends on the spin configuration
between site 1 and site 3.
When the spins of electrons on site 1 and site 3 are antiparallel,
the absolute value of the energy gain can be written approximately
as follows (see Fig.~\ref{sdproc}(a)),
\begin{equation}
  \frac{t_1^2}{u^{\prime}} + \frac{t_2^2}{u} 
+ \frac{t_3^2}{u'+\Delta_{e_g}}. 
\end{equation}
Here, $t_1$ represents the transfer between $\frac{1}{\sqrt{2}}(xy+yz)$
at site 1 and $\frac{1}{\sqrt{2}}(xy+yz)$ at site 3, 
and $t_2$ represents that between $\frac{1}{\sqrt{2}}(xy+yz)$
at site 1 and $\frac{1}{\sqrt{2}}(-xy+yz)$ at site 3, 
and $t_3$ represents that between $\frac{1}{\sqrt{2}}(xy+yz)$
at site 1 and $3z^2-r^2$ at site 3, 
and $\Delta_{e_g}$ denotes the level-energy difference
between $t_{2g}$ and $e_g$ level in the cubic crystal field.
On the other hand, when the spins are parallel, transfer to the 
$\frac{1}{\sqrt{2}}(-xy+yz)$
orbital is forbidden by Pauli's principle but the 
energies of the intermediate states in which two electrons occupy 
different orbitals are reduced by the intrasite exchange 
interaction $j$ (see Fig.~\ref{sdproc}(b)).
Consequently, the absolute value of the
energy gain can be written as 
\begin{eqnarray}
\frac{t_1^2}{u'-j}&+&\frac{t_3^2}{u'+\Delta_{e_g}-j}
\nonumber \\
&\sim&  \frac{t_1^2}{u'} 
+ \frac{t_1^2 j}{u'^2}
+ \frac{t_3^2}{u'+\Delta_{e_g}} 
+ \frac{t_3^2 j}{(u'+\Delta_{e_g})^2}. 
\end{eqnarray}
Therefore, the spin configuration between site 1 and site 3 is
determined by the competition between the following two energies,
$\frac{t_2^2}{u}$ and  
$\frac{t_1^2 j}{u'^2}+\frac{t_3^2 j}{(u'+\Delta_{e_g})^2}$.
In Fig.~\ref{egain_Jconbi} (upper panel), the values 
of these energies are plotted as functions of the bond angle.
As the ${\rm GdFeO}_3$-type distortion increases, 
the indirect hybridizations between neighboring $t_{2g}$ orbitals
are decreased and those between neighboring $t_{2g}$ orbitals and
$e_g$ orbitals are increased.
Consequently, the value of $t_2^2/u$ is decreased and that of
$t_3^2 j/(u^{\prime}+\Delta_{e_g})^2$ is increased, resulting in
the crossing of the two energies as the bond angle is decreased.
Moreover, since the hybridization between neighboring 
$e_g$ and $t_{2g}$ orbitals mediated by O $2p$ orbitals 
(i.e., $t_3$) becomes to have 
a $\sigma$-bonding character with the ${\rm GdFeO}_3$-type distortion,
the amplitudes of the $t_3$ and
$t_3^2 j/(u^{\prime}+\Delta_{e_g})^2$ are critically increased. 

On the basis of the above discussions, 
we can well describe this system by the
following Heisenberg model as long as the ($yz,zx,yz,zx$)-type
orbital order is strongly stabilized and hardly
affected by the change in the spin configuration,
\begin{equation}
    H_{\rm Heis} =
  J_{\rm Heis}^c \sum_{\langle i,j\rangle}^{c}    
{\vct{$S$}}_i\cdot{\vct{$S$}}_j 
 +J_{\rm Heis}^{a,b} \sum_{\langle i,j\rangle}^{a,b}
{\vct{$S$}}_i\cdot{\vct{$S$}}_j , 
\label{eqn:eqHeis}   
\end{equation}
with 
\begin{eqnarray}
  J_{\rm Heis}^{a,b} &<& 0 , \\
  J_{\rm Heis}^{c}
&=&4\left(\frac{t_2^2}{u}-
(\frac{t_1^2 j}{u'^2}+\frac{t_3^2 j}{(u'+\Delta_{e_g})^2})\right).   
\end{eqnarray}
Here, $\sum_{\langle i,j\rangle}^{c}$ denotes the summation
over the neighboring spin couplings along the $c$-axis and
$\sum_{\langle i,j\rangle}^{a,b}$ in the $ab$-plane.
The AFM(A)-to-FM2 phase transition occurs by the change 
in the sign of $J_{\rm Heis}^{c}$.
Moreover, within this model, 
the value of $J_{\rm Heis}^{c}$ decreases from
a positive to a negative value continuously 
as the bond angle is decreased and becomes 
zero at the phase boundary
as shown in Fig.~\ref{egain_Jconbi} (lower panel). 
Two-dimensional spin coupling
is realized at the phase boundary.
Consequently, the critical temperature 
at the phase boundary is suppressed to zero,
in accordance with  Mermin and Wagner's theorem~\cite{Mermin66}.  

However, the orbital states are actually slightly different between
AFM(A) and FM2 solutions.
In order to investigate how this slight difference in
orbital structures affects the spin-exchange interactions
along the $c$-axis and those in the $ab$-plane, we
estimate the values of $J_{\rm Heis}^{c}$ 
and $J_{\rm Heis}^{a,b}$ as functions of the bond angle
for the optimized orbital states of the stabilized spin configuration.
$J_{\rm Heis}^{c}$ and $J_{\rm Heis}^{a,b}$ 
are represented as,
\begin{eqnarray}
   J_{\rm Heis}^{c}
= (E_{\rm FM}-E_{\rm AFM(A)}) /2S^2,    \\
  2J_{\rm Heis}^{a,b}
= (E_{\rm FM}-E_{\rm AFM(C)}) /2S^2. 
\end{eqnarray}  
Here, $E_{\rm FM}$, $E_{\rm AFM(A)}$ and $E_{\rm AFM(C)}$ 
are the energy gains per unit formula
for FM, AFM(A) and AFM(C) spin configurations, respectively,
which are due to the second-order perturbational processes
under the orbital states of stabilized solutions.
\begin{figure}[tdp]
\hfil
\includegraphics[scale=0.5]{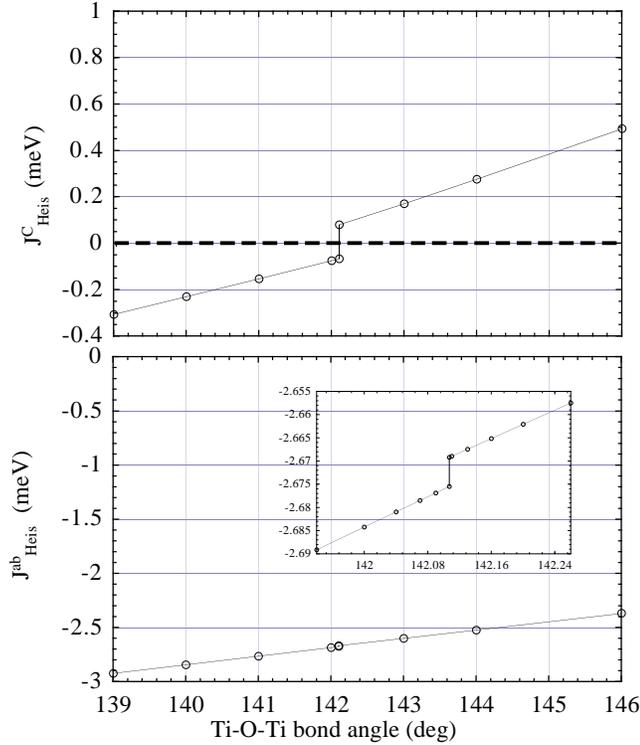}
\hfil
\caption{The value of the spin-exchange interaction along the $c$-axis
($J_{\rm Heis}^c$) and that in the $ab$-plane ($J_{\rm Heis}^{a,b}$)
are plotted as functions of the Ti-O-Ti bond angle.
While a small jump of the $J_{\rm Heis}^c$ value arises by a slight
change of the orbital state, the value is suppressed
to almost zero at the phase boundary. The inset of the lower panel
magnifies the small jump of the $J_{\rm Heis}^{a,b}$
at the phase boundary.}
\label{Jconbi}
\end{figure}
In Fig.~\ref{Jconbi}, the variations of   
$J_{\rm Heis}^{c}$ and $J_{\rm Heis}^{a,b}$
are plotted as functions of the bond angle.
While $J_{\rm Heis}^{a,b}$ constantly has a large negative
value ($\sim -2.5$ meV; ferromagnetic coupling), a positive value
of $J_{\rm Heis}^{c}$ in the relatively large bond-angle region
decreases as the bond angle decreases, and suppressed
to very small positive value as the Ti-O-Ti bond angle goes
to a transition point ($\sim 142.1^{\circ}$).
Then, $J_{\rm Heis}^{c}$ slightly jumps to a negative value, 
caused by a small change in the orbital state at the phase boundary.  
Although a small jump of the $J_{\rm Heis}^{c}$ 
($\Delta J_{\rm Heis}^{c}$) from a positive value ($\sim 0.1$ meV)
to a negative one ($\sim -0.1$ meV) occurs, 
the absolute values of $J_{\rm Heis}^{c}$ and 
$\Delta J_{\rm Heis}^{c}$ at the phase boundary
is sufficiently small relative to that of 
$J_{\rm Heis}^{a,b}$ ($\sim 2.7$ meV).
Therefore, we can conclude that the strong two-dimensionality
in spin couplings is realized near the phase boundary.
At this stage, the scenario of the rapid decrease
of $T_N$ is as follows:
The ${\rm GdFeO}_3$-type distortion increases the indirect hybridizations
between neighboring $t_{2g}$ and $e_g$ orbitals.
As a result, ($yz,zx,yz,zx$)-type orbital order
is strongly stabilized by the energy gain due to the transfers
almost independently of the spin configuration.
Since the orbital structure changes only little through
the magnetic phase transition, the spin-exchange interaction
along the $c$-axis decreases from a positive value to 
a negative one almost continuously
with increasing the ${\rm GdFeO}_3$-type distortion, and consequently  
becomes almost zero ($\sim0.1$ meV) at the phase boundary while that in the 
$ab$-plane remains ferromagnetic at $-2.67$ meV.
This strong two-dimensionality at the phase boundary 
suppresses the transition temperatures $T_N$ and
$T_C$.  

\begin{figure}[tdp]
\hfil
\includegraphics[scale=0.5]{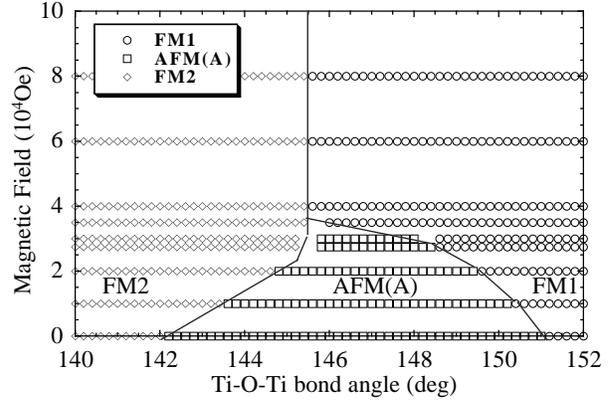}
\hfil
\caption{The magnetic phase diagram under the
application of the magnetic field.
The value of the JT-distortion-parameter
takes 1.030.}
\label{shift}
\end{figure}

In Fig.~\ref{shift}, we show the magnetic phase diagram
as a function of the bond angle
under the application of the magnetic field
in the spin-ordering direction ($z$-direction).
With increasing the magnetic field, the ground-state AFM(A) 
spin configuration changes into FM one at
a threshold value of the magnetic field.
The threshold value of the AFM(A)-to-FM2
spin-flip transition driven by the
application of the magnetic field
linearly decreases as the Ti-O-Ti bond angle
decreases.
This may reflect the linear behavior of $J_{\rm Heis}^c$ 
as a function of the bond angle.
Although we may expect more complicated behavior in the 
presence of the magnetic anisotropy in the compounds,
the qualitative structure of the phase diagram
in Fig.~\ref{shift} remain valid for small anisotropy.

\begin{figure}[tdp]
\hfil
\includegraphics[scale=0.5]{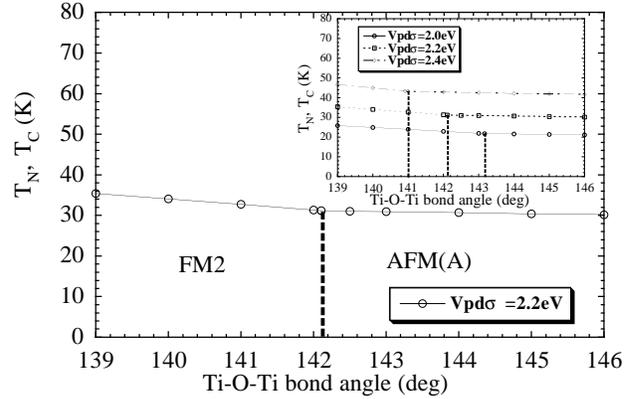}
\hfil
\caption{Magnetic transition temperatures $T_N$ 
and $T_C$ calculated by the mean-field 
approximation in the case $V_{pd\sigma}= 2.0$ eV 
are plotted as functions of the Ti-O-Ti bond angle.
The inset shows $T_N$ and $T_C$ in the case of 
$V_{pd\sigma}= 2.0$ eV, $2.2$ eV and $2.4$ eV.}
\label{tntc}
\end{figure}
In Fig.~\ref{tntc}, we show the values of 
$T_N$ and $T_C$ which are 
calculated by using the mean-field approximation.
A value of $T_C$ at $\angle$ Ti-O-Ti = $140^{\circ}$
is in good agreement with the experimental result for ${\rm YTiO}_3$.
However, the suppression of the
critical temperature at the FM-AFM phase boundary which is
expected to be realized from the two-dimensional character
cannot be reproduced within this mean-field calculation.
Instead, we can discuss the behavior of the critical temperature
qualitatively in the following manner:
Near the phase boundary, the spin coupling in this system
can be well described by the Heisenberg model given
in Eq. (\ref{eqn:eqHeis})
since the orbital state is hardly affected
by the spin state.
Besides, the strong spin-exchange interaction 
in the $ab$-plane $J_{\rm Heis}^{a,b}$ 
and the weak interaction along the $c$-axis $J_{\rm Heis}^{c}$ 
realize the strong two-dimensionality in the spin coupling;
\begin{equation}
     |J_{\rm Heis}^c| \ll |J_{\rm Heis}^{a,b}|.
\end{equation}
At this stage, it is justified to deal with the
spin-exchange term along the $c$-axis $H_{\rm Heis}^{c}$
within the mean-field approximation.
Then, we can write the magnetic susceptibility of this system
$\chi_{3D}$ as,
\begin{equation}
      \chi_{3D} = \frac{\chi_{2D}}{1-J_{\rm Heis}^{c}\chi_{2D}}.          
\end{equation}
Here, $\chi_{2D}$ denotes the magnetic susceptibility
of the two-dimensional spin system described 
by the Hamiltonian,
\begin{equation}
    H_{\rm Heis} =
  J_{\rm Heis}^{a,b} \sum_{\langle i,j\rangle}^{a,b}
{\vct{$S$}}_i\cdot{\vct{$S$}}_j , 
\end{equation}
Because of the strong spin-exchange interaction in the $ab$-plane,
when the system undergoes the magnetic phase transition,
the spin correlation in the $ab$-plane is expected to be sufficiently
large and the system is to be in the renormalized-classical
regime where $\chi_{2D}$ has a form,
\begin{equation}
      \chi_{2D} \propto \exp(\frac{J_{\rm Heis}^{a,b}}{T}).
\end{equation}
In addition, $J_{\rm Heis}^c$ shows a linear behavior as a function of 
the angle $\theta-\theta_{\rm c}$ near the phase boundary;
\begin{equation}
    J_{\rm Heis}^c \propto |\theta - \theta_{\rm c}|. 
\end{equation}
So that, the critical temperature near the phase boundary
which is defined by the divergence of $\chi_{3D}$ can be written
as a function of $\theta-\theta_{\rm c}$ as follows,
\begin{equation}
T_{\rm crit.}\propto\frac{1}{-{\rm ln}(|\theta-\theta_{\rm c}|)+const.}.
\end{equation}
This expression implies the rapid suppression of the critical temperature
as the system goes to the AFM-FM phase boundary.

Moreover, the large $T_N$ values ($\sim 120$ K) realized 
in the small ${\rm GdFeO}_3$-type distortion or the La-rich
region cannot be reproduced within this calculation.
We also calculated the magnetic transition temperatures in the
case $V_{pd\sigma}= 2.0$ and 2.4 eV (see Inset of Fig.~\ref{tntc}).
Although the values of $T_C$ and $T_N$
increases as the value of $V_{pd\sigma}$ is increased,
the large $T_N$ value observed in experiments cannot be reproduced.
This indicates that the AFM state realized in the
${\rm LaTiO}_3$ or in the less distorted region and that near the FM-AFM
phase boundary or in the more distorted region are qualitatively different.
In the AFM(A) or FM2 region, the ($yz,zx,yz,zx$)-type orbital
order is stabilized irrespective of the spin structure.
Consequently, the energy gain which depends on the spin structure 
may be rather small, resulting in the effectively small spin-exchange    
interaction and small $T_N$ and $T_C$ values.
\begin{figure}[tdp]
\hfil
\includegraphics[scale=0.5]{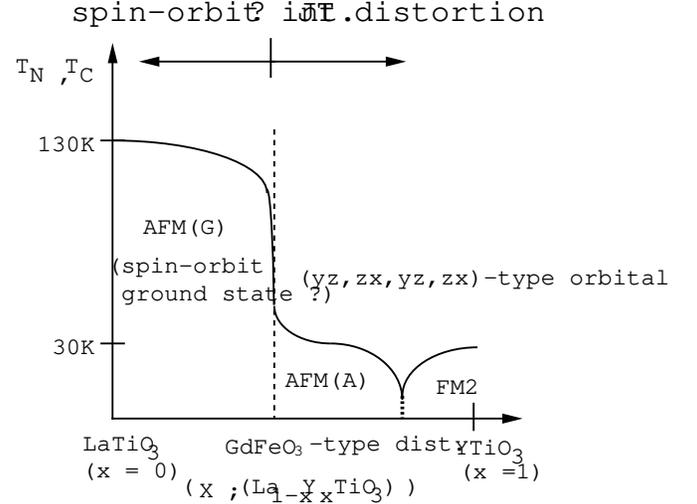}
\hfil
\caption{Schematic magnetic phase diagram.
$T_N$ takes relatively large
value in the AFM(G) phase realized in the 
region of the small ${\rm GdFeO}_3$-type
and $d$-type distortions.
With increasing the ${\rm GdFeO}_3$-type distortion,
the AFM(A) phase accompanied by the ($yz,zx,zx,yz$)-type
orbital ordering is realized with small AFM(A) spin coupling
along the $c$-axis. On the contrary, the rather strong FM coupling
is realized in the $ab$-plane and this drives the strong two dimensional
character in this system.
This causes the rapid suppression of the $T_N$.
The AFM spin-coupling along the $c$-axis is decreased almost 
linearly and strongly suppressed 
as system goes to the AFM(A)-to-FM2 phase boundary.
This causes the second-order like behavior of
this AFM to FM phase transition.
The JT distortion is relevant to the orbital state
in AFM(A) and FM2 phases while in the AFM(G) phase,
the spin-orbit interaction is considered to be substantial.} 
\label{schemag}
\end{figure}
On the contrary, the spin and orbital degrees of freedom 
in ${\rm LaTiO}_3$ strongly couple to each other,
which may result in the effectively strong spin-exchange 
interaction and large $T_N$ value. 
It may be necessary to take the spin-orbit interaction 
into consideration to reproduce the behavior of $T_N$
in the less distorted region.
According to the experimentally observed magnetic phase diagram,
$T_N$ rapidly decreases from rather large value
($\sim$ 100 K) around the distortion $\sim 151^{\circ}$. 
This suggests that the AFM spin coupling along the $c$-axis
is already suppressed strongly
when the AFM(A) phase accompanied by 
the ($yz,zx,yz,zx$)-type orbital state
is stabilized with increasing the magnitude of the ${\rm GdFeO}_3$-type
distortion.  
Consequently, the schematic magnetic phase diagram
can be obtained as shown in Fig.~\ref{schemag}.

Moreover, the critical angle decreases as the value of
$V_{pd\sigma}$ increases.
This behavior seems as if the large $d$-$d$ transfer
would not favor the FM2 state.
With increasing the $V_{pd\sigma}$ value, the energy difference 
between the $t_{2g}$ and $e_g$ levels
(i.e., $\Delta_{e_g}$) are increased, which is 
approximately given in as Eq. (\ref{eqn:eqdleg}).
This increase of $\Delta_{e_g}$ strongly suppresses the
characteristic perturbational 
energy gain for the FM2 spin configuration
$t_3^2j/(u'+\Delta_{e_g})^2$. 
Consequently, the AFM phase tends to be enhanced 
while the FM phase suppressed with a large $V_{pd\sigma}$ value.

%% file: sec4.tex
\section{Summary, Conclusions and Discussions}

In this paper, we have studied the magnetic and orbital ordered states
and their phase transitions in perovskite-type Ti oxides
by utilizing the effective spin and pseudospin Hamiltonian.
The Hamiltonian is derived by the second-order
perturbational expansion excluding the doubly occupancy of the ${t_{2g}}$
states in the limit of the strong Coulomb repulsion.
The full degeneracies of Ti $3d$ and O $2p$ orbitals 
and $d$-$d$ Coulomb and exchange interactions are
taken into account.
Moreover, the effects of the ${\rm GdFeO}_3$-type distortion
and the $d$-type JT distortion are also considered through
the anisotropic transfer amplitudes and the Ti $3d$ 
level splitting.
We neglect the spin-orbit interaction in the region of large
JT distortion. 
According to the mean-field calculations, in the $d$-type JT distortion,
the ($yz,zx,yz,zx$)-type orbital order is strongly favored 
by the ${\rm GdFeO}_3$-type distortion
due to the large amplitudes of transfers between neighboring $e_g$ 
and $t_{2g}$ orbitals.
The AFM(A) phase with this orbital ordering 
is stabilized in the moderately distorted region.
This AFM(A) solution is obtained theoretically for the first time
in the perovskite-type Ti oxides as far as we know. 
This phase may be detected by the neutron-scattering experiments.
 
With decreasing the Ti-O-Ti bond angle, the AFM(A)-to-FM
phase transition arises.
The $e_g$-orbital degrees of freedom play an important 
role on this magnetic phase transition.
The energy difference between AFM(A) and FM solutions
is consistent with the value expected from 
experimentally observed $T_N$ and $T_C$.

The ${\rm GdFeO}_3$-type distortion makes the $t_{2g}$ orbitals
hybridized with the $e_g$ orbitals.
In the $d$-type JT distortion, ($yz,zx,yz,zx$)-type orbital state is
stabilized due to the large amplitudes of the transfers
between neighboring $t_{2g}$ and $e_g$ orbitals.
Through the AFM(A)-to-FM phase transition, this 
orbital state hardly changes.
Since the orbital state changes negligibly through  
the magnetic phase transition,  the spin-exchange interaction
along the $c$-axis which is characterized by the 
second-order perturbational energy gains
changes from positive (AFM spin coupling) to
negative (FM spin coupling) nearly continuously
and is suppressed to almost zero at the phase boundary.
On the contrary, the strong FM spin coupling 
is constantly realized in the $ab$-plane.
Consequently, the strong two-dimensionality in the 
spin coupling is realized near the phase boundary.
This fact is a possible reason for the
critical suppression of $T_N$ and $T_C$
near the phase boundary.
We expect strong quantum fluctuations at the phase boundary.
A slight change in the orbital state occurs at the phase boundary, and
it causes a very weak first-order phase transition in this system.

This two-dimensional character in the spin coupling may
be reflected on the dispersion relations of the spin waves
which can be obtained by neutron-scattering experiments. 
When the spin coupling in this system is mapped on the
Heisenberg model given in Eq. (\ref{eqn:eqHeis}),
the dispersion relation of the spin wave in the AFM(A) state
($J_{\rm Heis}^{a,b} < 0$, $J_{\rm Heis}^{c} > 0$) is given
by 
\begin{eqnarray}
&&\eps(\vct{$k$}) = \nonumber \\
&&\!\!\!\!\!\!\!\!\!\!\sqrt{\![\!-\!J_{\rm Heis}^{a,b}\!\{2\!\!-\!\!
(\!\cos{k_xa}\!\!+\!\!\cos{k_ya}\!) 
\!\}\!\!+\!\!J_{\rm Heis}^{c} ]^2\!\!\!-\!\!
(\!J_{\rm Heis}^{c}\!\!\cos{k_za})^2  },
\end{eqnarray}
where $a$ is a lattice constant.
This expression can be easily obtained by 
the conventional Holstein-Primakoff transformation. 
We can rewrite this dispersion relation  
near the ($k_x, k_y, k_z) = (0,0,0)$ point as,
\begin{equation}
     \eps(\vct{$k$}) = \sqrt{-J_{\rm Heis}^{a,b} J_{\rm Heis}^c
          (k_x^2a^2 + k_y^2a^2)  +
         {J_{\rm Heis}^c}^2 k_z^2a^2}.
\end{equation}
If $|J_{\rm Heis}^{c}| \ll |J_{\rm Heis}^{a,b}|$ is realized,
the ratio for the slope of the dispersion in the $z$-direction
to that in the $x,y$ directions 
(i.e., $\sqrt{|{J_{\rm Heis}^{c}}^2/
J_{\rm Heis}^{a,b}J_{\rm Heis}^{c}|}
= \sqrt{|J_{\rm Heis}^{c}/J_{\rm Heis}^{a,b}|}$) will be suppressed 
strongly.
In this way, the strong two-dimensionality in the spin coupling 
near the phase boundary which is predicted in our 
theory will be reflected on the dispersion.
Besides, we can expect to observe 
experimentally:
\begin{itemize}
\item in AFM and FM phases near the phase boundary, 
or in ${\rm SmTiO}_3$ and ${\rm GdTiO}_3$,
the orbital states are similar to that in ${\rm YTiO}_3$.
\item AFM state near the phase bounadary is A-type.
\item Strong quantum fluctuations due to the two dimensionality
exist near the phase boundary. 
\end{itemize}
Moreover,we can control the dimensionality by $R$-site substitution
in this system.  

Similar effective spin and pseudospin Hamiltonians
have also been proposed for perovskite-type Mn oxides 
in which not only diagonal transfers but also
off-diagonal transfers between neighboring $e_g$ orbitals
are taken into account
~\cite{Kugel72,Kugel73,Kugel82,Castellani78,Koshibae97,Ishihara96,Ishihara97}.
In those Hamiltonians, the transfer integrals
reflect the realistic cubic perovskite-type structure.
However, the effects of the ${\rm GdFeO}_3$-type distortion
are not introduced in those models. 
It should be noted that the hybridization between $t_{2g}$ and $e_g$
orbitals driven by the ${\rm GdFeO}_3$-type distortion
is of crucial importance for the stabilizing
the AFM(A) and FM2 phases in perovskite-type
Ti oxides, as we have discussed.

In connection to the double exchange mechanism, 
we point out the possibility of an interesting
large negative magnetoresistance around the AFM(A)-to-FM2
transition point when carriers are doped by substitution 
of the $R$-site with an element with different valence such as Sr.
Since the $c$-axis conduction is strongly favored by the ferromagnetic
ordering, the applied magnetic field 
in the AFM(A) phase near the transition point may contribute to 
drastically reduce the resistivity and favors the appearance of a 
metal under the carrier doping.

\section*{Acknowledgement}
M. M. thanks T. Mizokawa and H. Asakawa
for valuable discussions and useful comments. 
This work is supported by ``Research for the Future Program''
(JSPS-RFTF97P01103) from the Japan Society for the Promotion of Science.